\documentclass[12pt,preprint]{aastex}

\received{2006 August 4}
\begin{document}

\title{VLA Observations of J1228+441, a Luminous Supernova Remnant  in
NGC 4449 }

\author{Christina K. Lacey} \affil{University of South Carolina,
Department of Physics and Astronomy, Columbia, SC 29208}
\email{lacey@sc.edu}

\author{W. M. Goss} \affil{ P.~O.~Box 0, National Radio Astronomy
Observatory, Socorro, NM 87801} \email{mgoss@nrao.edu}

\author{Leila K. Mizouni} \affil{University of South Carolina,
Department of Physics and Astronomy, Columbia, SC 29208}
\email{mizouni@physics.sc.edu}

\begin{abstract}

The luminous, oxygen-rich supernova remnant, J1228+441, is located in
  the irregular galaxy NGC 4449 and has  been observed at radio
  wavelengths for thirty years. An analysis of recent VLA observations
  of NGC 4449,  combined with   VLA archive data and previously
  published VLA and WSRT  observations, yields light curves at 6 and
  20 cm from 1972 to 2002.  
     The light curves at all radio frequencies
  exhibit a marked decline in radio emission, confirming past findings.      This paper presents and
  discusses the radio light curves and spectral index, $\alpha$,
  variations  from 1972 to 2002 where $S_\nu \propto \nu^{-\alpha}$ and compares J1228+441 with other radio supernovae. 
  The spectral index of J1228+441 
  appears to have steepened in the last five years
  at higher frequencies from $\alpha=0.64\pm 0.02$ in 1996 to
  $\alpha=1.01\pm 0.02$ in 2001-2002.

 \end{abstract}
 

\section{Introduction \label{intro}}

 NGC 4449 is a Magellanic type irregular galaxy with several areas of
vigorous recent star formation, including the nucleus and the bar.
The distance to NGC 4449 has been suggested to range from  3.7 Mpc
(Bajaja et al. 1994) to 5.0 Mpc (Sandage \& Tammann 1975 and Aaronson
\& Mould 1983). In this paper, we adopt a distance of 3.9 Mpc (de
Vaucouleurs 1975).  The recent star formation in NGC 4449 may have
been triggered by a past galactic collision and tidal stripping
(Hunter, van Woerden, \& Gallagher 1999).  The massive stars formed in
the recent star formation episodes can then become the progenitors for
core-collapse supernovae.

 The ultra-luminous supernova remnant (SNR), J1228+441, was first detected by Seaquist \&
Bignell (1978) and Balick \& Heckman (1978) at radio and optical
wavelengths, respectively.
 The nonthermal spectral index, $\alpha$=0.95 $\pm$0.27  ( $S_\nu
\propto \nu^{-\alpha}$, where $S$ is the flux density at frequency
$\nu$) (Seaquist \& Bignell 1978), and observations of both broad and
narrow lines in the optical spectrum (Balick \& Heckman 1978) of
J1228+441 indicate that the source was a SNR.   Additional optical
spectroscopic observations by Kirshner \& Blair (1980), Blair,
Kirsher, \& Winkler (1983)  confirmed that  the SNR  is embedded in
and interacting with a  nearby \ion{H}{2}  region.

An upper limit of the diameter of J1228+441  was determined first by
De Bruyn (1983) who used European VLBI data to place a limit of $\le
0\farcs07$ ( 1.3 pc at 3.9 Mpc) on the  angular size of J1228+441.
The age of J1228+441 has been estimated from  expansion velocities  in
excess of 6,000 km~s$^{-1}$, based on  UV/optical HST and ground based
observations (Blair \& Fesen 1998, Blair 2006).   The age of J1228+441
was estimated to lie in the range of 60 - 200 years.  In this paper,
we adopt an explosion date of 1900 for J1228+441 with a large
uncertainty.  Blair \& Fesen (1998) and Blair (2006)  found  an upper
limit of 0\farcs028 (or 0.5 pc) for the diameter of the SNR based on
HST data. J1228+441 is  one of the few known intermediate aged SNRs
with an age  between that of Cas A,  which is  $\sim$330 years (Thorstensen, Fesen, \& van den Bergh 2001), and that of
the  oldest known extragalactic radio   supernovae, SN 1923A in M83
(Eck et al. 1998), which is 83 years old, and SN 1957D also in M83 (Pennington \& Dufour 1983), which is 49 years old.
J1228+441  is also notable as the most luminous and most distant
member of the class of oxygen-rich SNRs (Blair, Kirshner, \& Winkler
1983).

 Radio observations of J1228+441 were   reported by de Bruyn, Goss, \&
 van Woerden (1981), de Bruyn (1983), and  Bignell and Seaquist
 (1983).   De Bruyn, Goss,  \& van Woerden (1981)  used Westerbork
 Synthesis Radio Telescope (WSRT) observations of J1228+441 and found
 a long-term trend of decline in the radio emission at 6 cm over a
 period of seven years.
Bignell and Seaquist (1983)  reported NRAO\footnote{The National Radio
 Astronomy Observatory is a facility of the National Science
 Foundation operated under cooperative agreement by Associated
 Universities, Inc.} Very Large Array (VLA) observations of J1228+441
 from 1979 to 1982
and concluded that the new observations showed significant variability
 at 5GHz. However, they could not establish the nature of the
 variability based on their data.  Additional WSRT observations from
 late 1980 to 1982  were reported by de Bruyn (1983), who found that
 the flux density of J1228+441 at 5 GHz confirmed the decline   first
 reported in  de Bruyn, Goss, \& van Woerden (1981).   
There is also weak evidence that the  X-ray emission from J1228+441
has decreased by roughly a factor of two from the {\it Einstein} data
of 1983 (Fabbiano et al. 1992)  compared to both the ROSAT data of
1997 (Vogler \& Pietsch  1997) and the {\it Chandra}  data of 2002
(Summers et al. 2003, Patnaude \& Fesen 2003).

\section{The VLA Observations of J1228+441}

\begin{figure}
\plotone{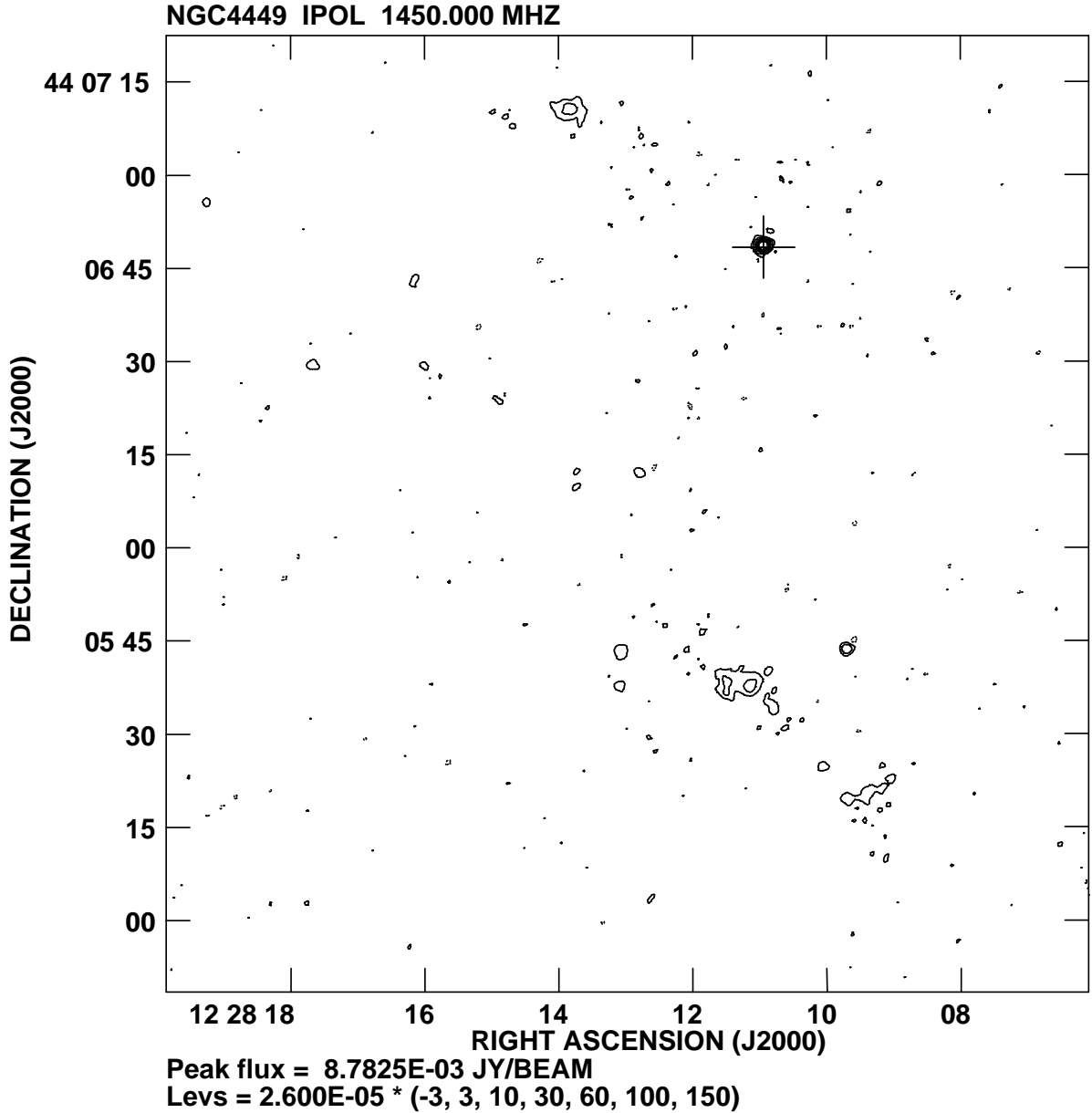}
\caption{The 20 cm emission from NGC 4449 is shown with contour levels
 of -3, 3, 10, 30, 60, 100, and 150 times 26 $\mu$Jy beam$^{-1}$.
 J1228+441, the luminous SNR, is the brightest source in NGC 4449 at
 20 cm and is  denoted by a cross. The image has been corrected for
 primary beam attenuation.    The observation was carried out on 30
 March 1994 and the observing parameters were: beam size = 1.3\arcsec
 $\times$ 1.2\arcsec, PA= 86\arcdeg,  and a rms noise of 26
 $\mu$Jy~beam$^{-1}$.     \label{snrkntr}}
\end{figure}

A series of VLA observations was undertaken in 1994; A array
observations at 20 cm with a beam of 1.3\arcsec $\times$ 1.2\arcsec,
PA= 86\arcdeg,  and a rms noise of 0.026 mJy~beam$^{-1}$ were carried
out on 30 March 1994, followed by B array observations at 6 cm with a
beam of 2.42\arcsec $\times$ 1.39\arcsec, PA= 82\arcdeg,  and a rms
noise of 0.028 mJy~beam$^{-1}$, 16-17 June 1994.  These observations
form the basis of a comprehensive study of NGC 4449, which will be
reported elsewhere (Lacey et al. 2006).  In  Figure \ref{snrkntr}, we
show the location of J1228+441 within the galaxy NGC 4449.  The source
is located 30\arcsec~ (570 pc, projected) north of the center
(position given by Hunter et al. 1998) of NGC 4449.

To follow the variations in flux density  in the intervening years
 since  de Bruyn (1983)  and Bignell \& Seaquist (1983)   the VLA
 archive was searched for observations of NGC 4449 and J1228+441 that
 had not been previously published.   A total of twelve 20 cm
 observations, fifteen 6 cm observations, and six additional
 wavelength observations were located in the archive. These
 observations  of J1228+441  are presented in Table~\ref{taball},
 containing the observation date, telescope, beam parameters, VLA
 configuration, frequency (20 cm ~1.4 GHz  and 6 cm ~4.8 GHz), flux
 density, and one sigma errors.     The WSRT data from de Bruyn
 (1983), the published VLA data from Seaquist \& Bignell (1978),
 Bignell \& Seaquist (1983), and Johnson et al. 2006  are also
 included in Table~\ref{taball}.  The WSRT data (Table~\ref{taball})
 are taken from only de Bruyn (1983), in which the WSRT data were
 corrected to the Baars et al. (1977) flux density scale. In addition,
 the observations from de Bruyn et al. (1981)  were corrected for
 radio confusion arising from  a nearby \ion{H}{2} region. De Bruyn
 (1983) notes that  the confusion in the  7\arcsec\ WSRT beam at 6  cm
 is  $<0.5$ mJy.  For the lower frequency VLA data (1.4 to 5 GHz), the
 effects of a confusing source of 380 mJy at 20 cm located 8\arcmin~
 to the southwest of J1228+441 must be subtracted in the imaging
 process. The confusing source is located at $\alpha(2000): 12^h 27^m
 42\fs00 \pm 0\fs02, \delta(2000): 44\arcdeg 00\arcmin 42\farcs0 \pm
 0\farcs2$.  For the spectral line observations at \ion{H}{1} (1.420
 GHz) and OH (1.720) GHz  only the  continuum channels were used.

\section{Time Dependent Decline of Radio Emission}

\begin{figure}
\plottwo{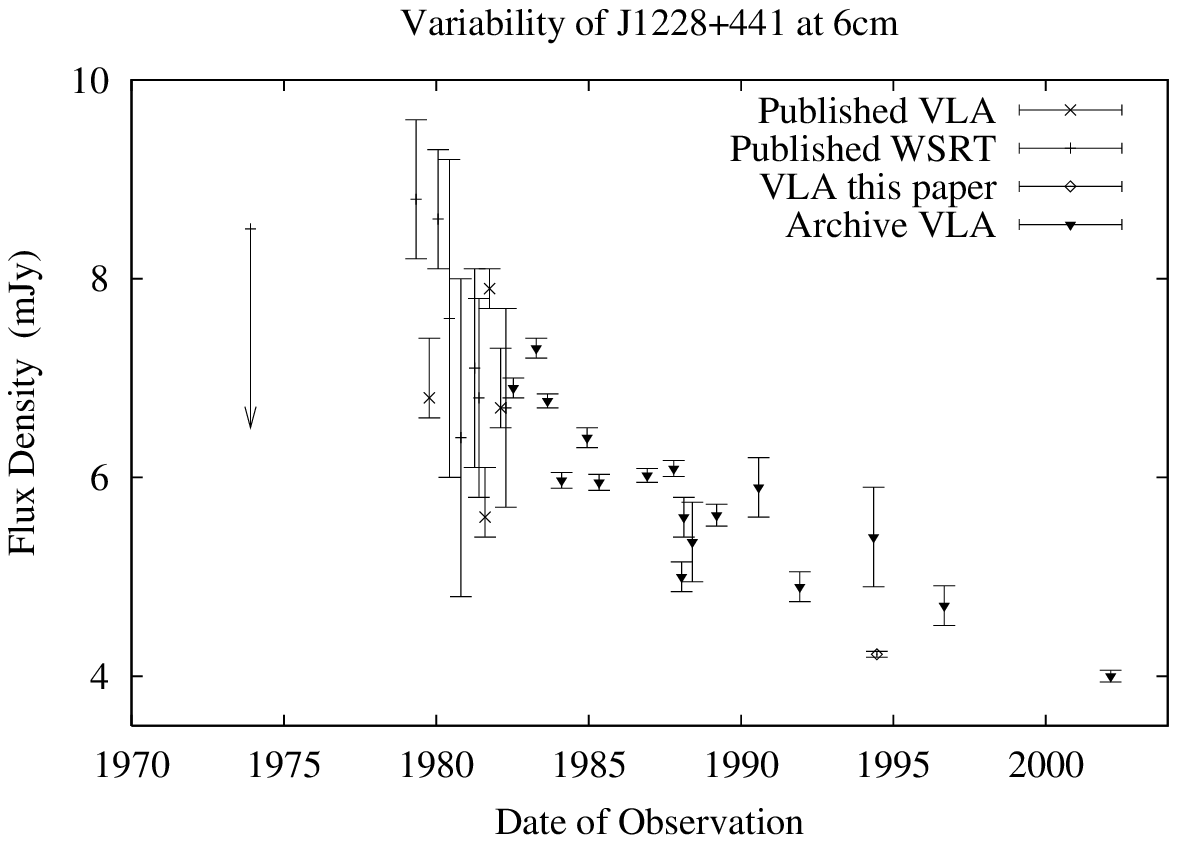}{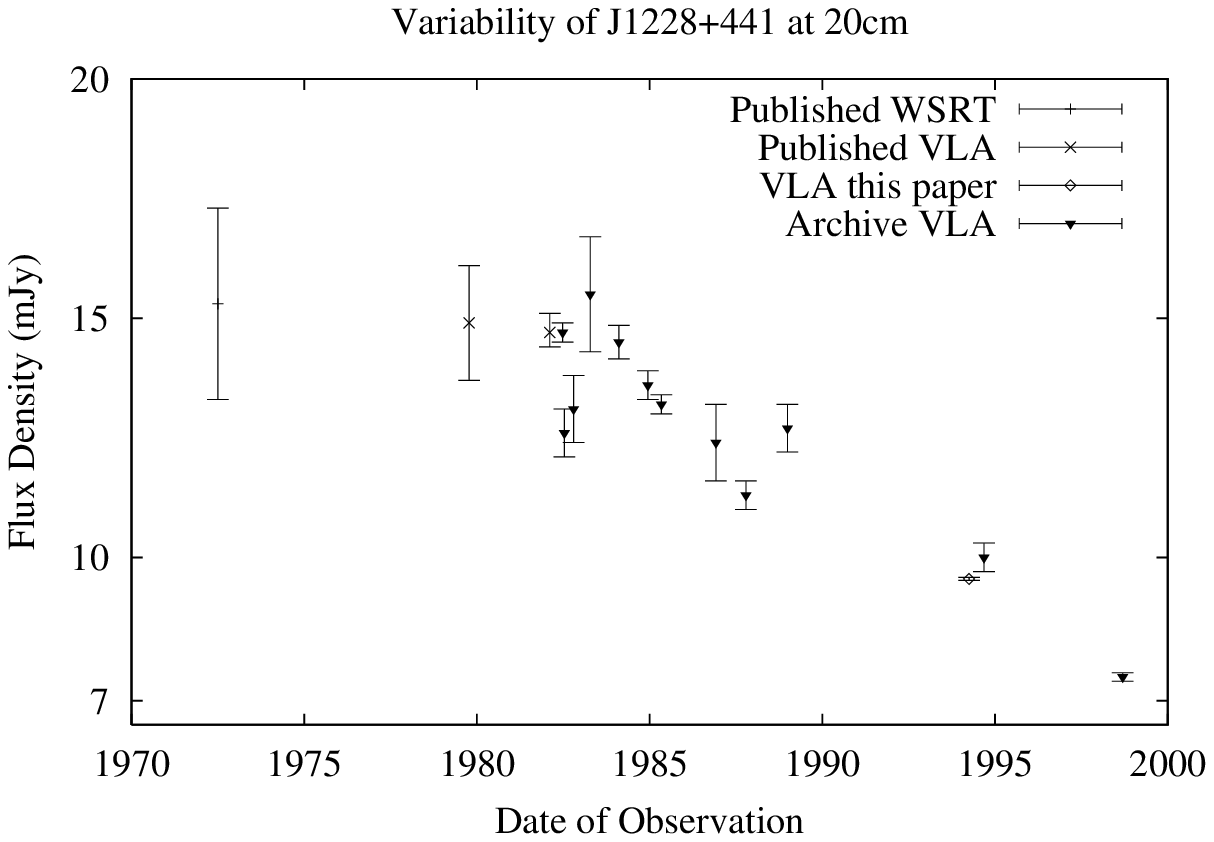}
\caption{The lightcurves of J1228+441 at 6 cm and 20 cm.  Both light
curves show a  significant decrease in the period of 1979 to 2002.
\label{lightcurve}}
\end{figure}

The flux density of J1228+441 has declined gradually over 30 years
(1972 to 2002).  The 20 cm and 6 cm light curves presented  in
Figure~\ref{lightcurve} clearly demonstrate a  decline in the radio
emission at both frequencies.  De Bruyn (1983) suggested that, in
addition to the smooth decline of radio emission, there were possible
small  variations from a smooth decline.   The VLA data at  6 cm
presented in this paper suggest that variations of 0.7$\pm$0.3 mJy
over short time scales (two to three years) may be present in addition
to the   overall  smooth decline,    but  the evidence for the
short-time scale  deviations from a smooth decline is only at a level
of 2$\sigma$.

\section{Evolution of the Spectral Index}

  De Bruyn, Goss,  \& van Woerden (1981)  compared their 5GHz
 measurements with Seaquist and Bignell (1978)  and found that
 J1228+441 must have undergone significant variability  in order to
 explain the variability in spectral index  measurements.   De Bruyn,
 Goss,  \& van Woerden (1981)  speculated that  the spectral index
 varied with a timescale of  a few years.  Bignell \& Seaquist (1983)
 also found that the spectral index must have changed from the
 earliest value of   $\alpha=0.98$ to   0.62, which was flatter.

\begin{figure}
\plotone{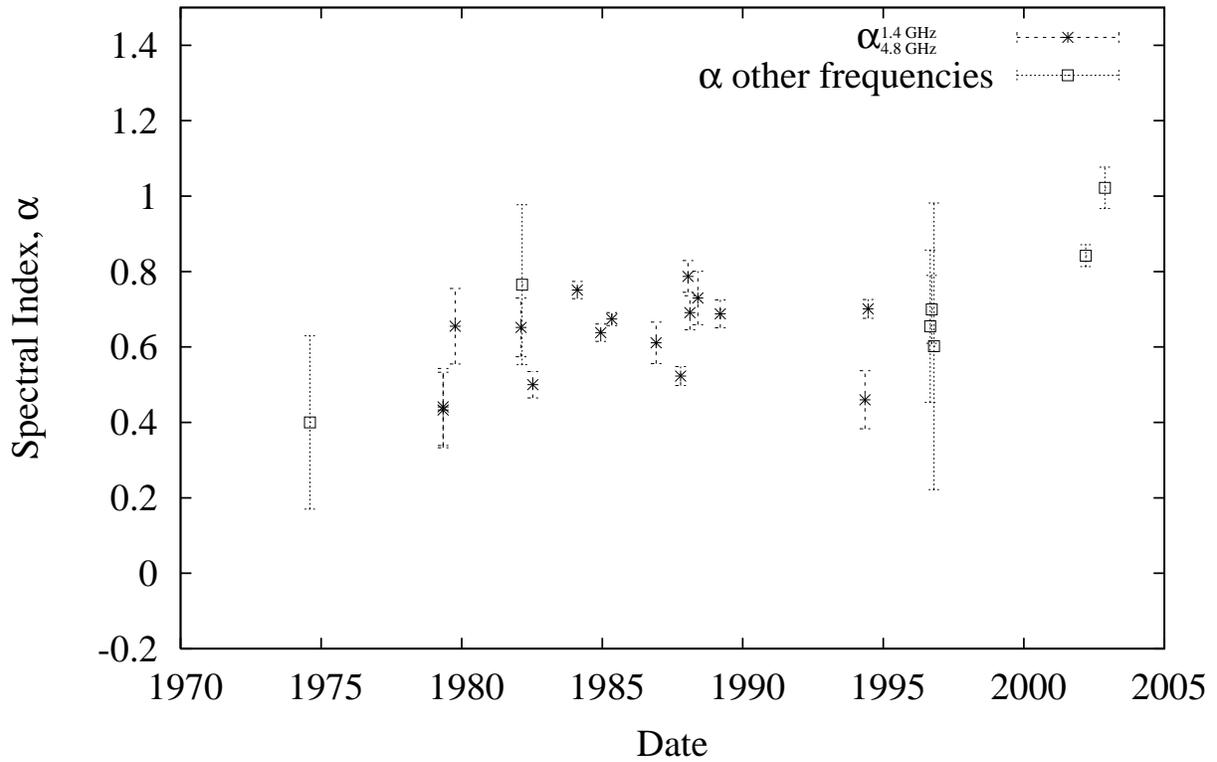}
\caption{Evolution of the spectral index from 1974 to 2002.
 \label{alphat}}
\end{figure}

Table~\ref{alpha} lists the dates  and the spectral index, $S_\nu
\propto \nu^{-\alpha}$, based on two frequencies that were observed
within an interval of 8 - 16 months.  Figure ~\ref{alphat} displays the
spectral index calculated from two frequencies for  J1228+441  over
the time period 1972 to 2002. The spectral index  is variable from
1972 until 1982 as noted by De Bruyn, Goss,  \& van Woerden (1981).
This early period is characterized by the largest flux density errors
and the 6 cm (4.8 GHz) flux density varies from $13.0 \pm 2.0 $ mJy in
1973, to an upper limit of $< 8.5$ mJy in 1973, to $10.9 \pm 2.9, 2.1$
mJy in 1974.   The spectral index  has a relatively
constant value of $\alpha=0.65 \pm 0.02$ from 1982 to 1996.
 This
period has high sensitivity VLA data and the 6 and 20 cm were observed on the same day, which eliminates temporal effects from the determination of the spectral index.   The spectral index then
steepens  to 1.0 $\pm$ 0.02 in 2001 -- 2002.  In 2001-- 2002, a 22 GHz
and  a 4.9 GHz frequency
observation  provided the most reliable measure of the spectral index since the frequency range covers an order of
magnitude.     In
summary, analysis of Figure ~\ref{alphat} shows an essentially
constant value from 1980 through 1997, with  possible  small scale
variations at the 1 -- 2~$\sigma$ level with a period of  two  to three
years.  After 1997, the spectral index steepens.

\begin{figure}
\plotone{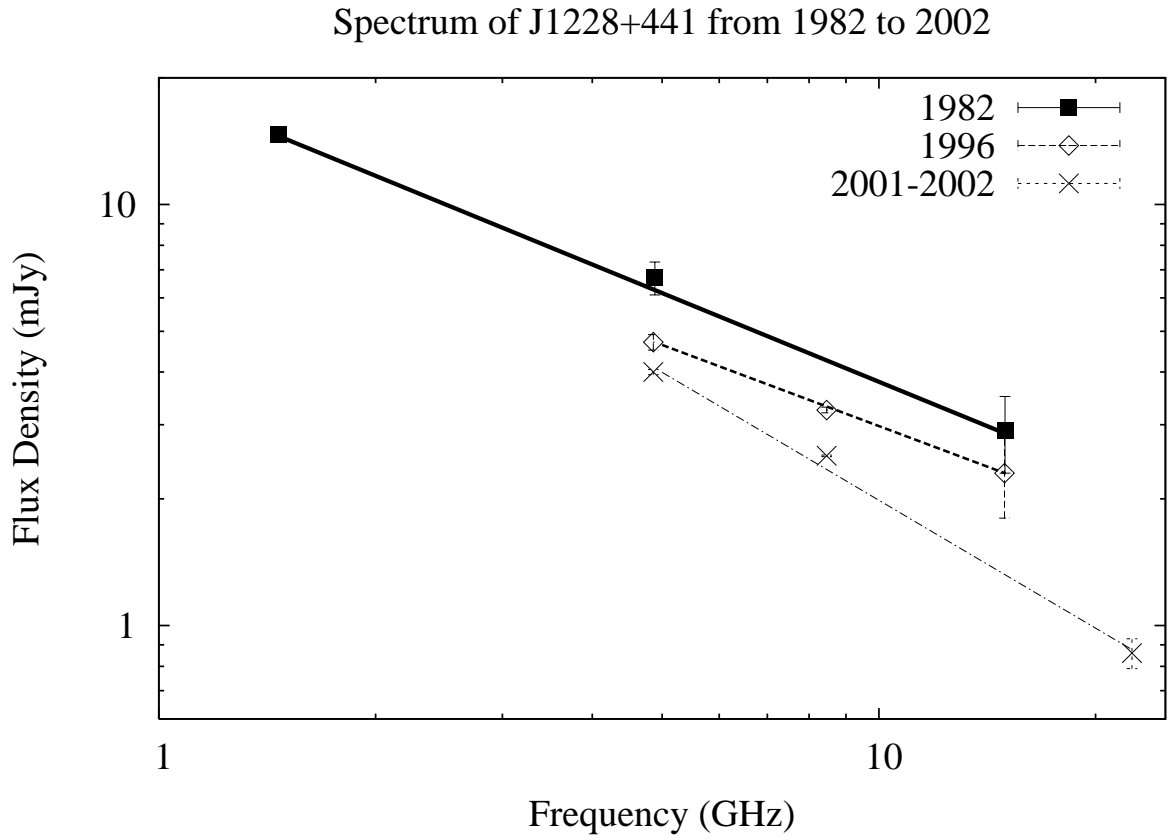}
\caption{The continuum spectra of J1228+441 from 1982 - 2002.   The
  points in each epoch were fitted to $S_\nu \propto \nu^{-\alpha}$,
  where $\alpha$ is the spectral index.   The values of $\alpha$ are:
  $\alpha=0.70\pm0.03$ in 1982, $\alpha=0.64\pm0.02$ in 1996, and
  $\alpha=1.0\pm0.02$ in 2001-2002.
  \label{specall}}
\end{figure}

 The spectrum of J1228+441 is shown in Figure~\ref{specall}   at three
epochs: 1982, 1996, and 2001-2002.  The fitted values of $\alpha$
shown in Figure~\ref{specall}  are:  $\alpha=0.70\pm0.03$ in 1982,
$\alpha=0.64\pm0.02$ in 1996, and $\alpha=1.0\pm0.02$ in 2001-2002.
  The spectrum of J1228+441 is only shown for cases in which three or
 more observations at substantially different frequencies were
 available within a time range of 
  8 months  for the spectra in 1994 and 1996 and
 within a time range of 16 months for the 2001-2002 spectrum.  As expected,  the early
 spectra exhibit a higher flux density for each  frequency than the
 later epoch spectra.  In addition, the 2001-2002 epoch spectrum
 displays  a steepening that is also reflected in the steepening of
 the spectral index (see Figure ~\ref{alphat}), providing additional
 support that the spectral index is steepening and that the steepening
 is not just due to the time difference between the observations in
 2001-2002.  

 Montes et al. (2000) also reported spectral index variations for the radio SN 1979C (see Section~\ref{secsne} for a discussion of the differences between radio SNe and SNRs); the spectral index both steepened and flattened over time scales on the order of  $\le$ 365 days   for the time period between 4300 -- 7100 days after the explosion. Unfortunately, a scarcity of high frequency observations and adequate time coverage of the radio emission from  SN 1979C during this period prevented the testing of hypotheses concerning the mechanism responsible for  the spectral index variations.     Both 
J1228+441 and SN 1979C highlight the need for future  observations of both objects over closely spaced time  intervals   covering
the full radio frequency range  in order to fully confirm the magnitude and time scale of the small-scale variations. 

 The steepening of the spectrum of J1228+441 is most likely caused by synchrotron or inverse Compton  radiation losses, possibly augmented by the effects of adiabatic expansion.  
  Extensive multi-wavelength monitoring of 
  J1228+331 over an extended period of time is required in order to produce spectra with enough precision to confirm the spectral steepening and identify the time period over which the spectral index changes. More precise determinations of the radio spectra are required in order
to determine the mechanism that is responsible for the likely spectral
steepening.

\section{Discussion \label{disc}}
 
 The fractional decline rate, $\frac{dS}{S dt}$, can be calculated for both the 6 cm and
 20 cm data from 1975 to 2001 using the relation.
   We find rates:    $\frac{dS}{S dt}=
 -2.3\%\pm$ 0.4\% per year  at 6 cm and  $\frac{dS}{S dt}= -2.8\%\pm$
 0.6\% per year at 20 cm.
The high
 sensitivity VLA
 observations dominate the estimate of the fractional decline rate.
  The decline  of J1228+441 is well fit by a constant fractional
 decline rate; there is no evidence that this 
 rate has changed over the $\sim$25 years of observations.
  These rates are noticeably less than the rate of $\sim   -6\%$ per
 year at 6 cm found by  de Bruyn, Goss, \& van Woerden (1981) for a
 shorter interval of 1975 to 1981.  
  The decline rate of J1228+441 is similar to that of Cas A,  which is
-1.3\% per year for the radio ring (Tuffs 1986) for the period of 1974
to 1978,  and much larger than that of the Crab Nebula, which is -0.17
\%$\pm$ 0.02 \% per year for the period of 1968 to 1984 (Aller \&
Reynolds 1985).
 
  If the flux density evolution is described by  $S(t)=
 S_0\left ( \frac{t}{t_0} \right )^{\beta}$, where $t$ is the time since the explosion, then $\beta$ is just  $\frac{dS}{S dt}$ multiplied by the age of the remnant (de Bruyn, Goss, \& van Woerden 1981).  Since the age of J1228+441 is not well known, we give $\beta$ at 6 cm  for a range of ages:  for an age of t=100 yrs, $\beta=-2.3$; for an age of 150 years, $\beta=-3.5$; and for an age of 200 yrs, $\beta=-4.6$.  Similarly, for 20 cm we find:  for an age of t=100 yrs, $\beta=-2.8$; for an age of 150 years, $\beta=-4.2$; and for an age of 200 yrs, $\beta=-5.6$.  The values of $\beta$ calculated for different ages of J1228+441 vary significantly due to the large error in the age and, thus, are not useful without a better estimate of the age.

\section{J1228+441 Compared with the Oldest Radio SNe \label{secsne}}

There is some ambiguity in the classification of radio supernovae and SNRs.    Supernovae are objects that are generally discovered at optical wavelengths soon after the initial explosion.  The explosion date of a SN is often known very precisely with current SN monitoring programs and all supernovae are  characterized by a light curve that over time exhibits an initial rise in the emission, a peak, and then a gradual  decline.  Radio emission from optically identified SNe  arises from the interaction of the supernova shock with the circumstellar medium (CSM) that was initially created by the progenitor star.  The CSM originates from the stellar winds of the red supergiant phase of the progenitor star and  typically expands with speeds around 10 km s$^{-1}$, which allows the much faster supernova shock to overrun the CSM.  The rise and decline of radio emission due to the CSM interaction with the shock has been modeled (Weiler et al. 2002 and references therein).  Radio SNRs are primarily objects that have ages of 300 -- 10,000 years and a typical radio SNR produces radio emission arising from the interaction of the shock with plowed up interstellar material. The youngest radio SNRs, e.g. the Crab Nebula and Cas A, show small declines in radio emission of a few percent per year.  Very few SNRs have known explosion dates; the few that are known, such as the Crab Nebula, are known due to historical records and measurements of shock expansion speeds. In the Milky Way Galaxy, a supernova has not occurred in the last 200 years and the youngest Galactic SNR, Cas A, exploded in the late 17th century with some uncertainty as to the exact explosion date (Thorstensen, Fesen, \& van den Bergh 2001).   The currently known supernovae are all extragalactic in origin.   The transition between the CSM dominated phase of a radio SN to the ISM dominated phase of a radio SNR has not been observed and the transition is predicted to depend on the CSM and the ISM distributions.  The environment, in particular the density and distribution of the ISM, of a SNR greatly affects the ability of a SNR to produce radio emission (Lacey and Duric 2001).
  
SN 1957D and SN 1950B in M83  and SN 1970G in M101 are  radio
supernovae whose radio emission has been identified and monitored for
at least three decades. SN 1961V in NGC 1058 was initially identified
as a peculiar SN, but the exact nature of SN 1961V is still uncertain
(Chu et al. 2004).     We  compared J1228+441 with the young Galactic SNRs, the Crab and Cas A, in Section~\ref{disc}.   Since J1228+441 has declining radio emission,  it is natural to compare this object with other older radio supernovae.     Table~\ref{radiosne} shows the  basic properties
of J1228+441 compared    with  SN 1957D, SN 1950B, and SN 1970G.

\begin{figure}
\plotone{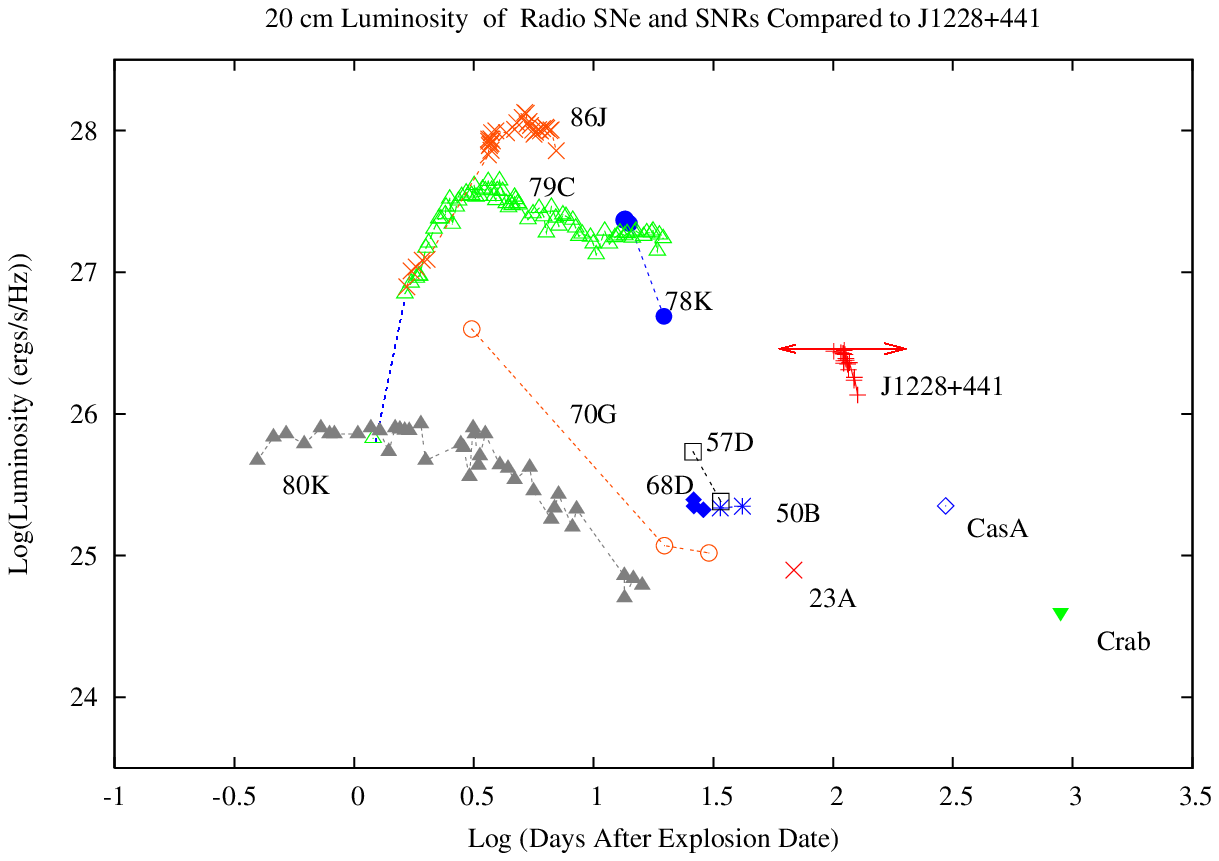}
\caption{A comparison of the luminosity at 20 cm and age of J1228+441
  with other radio supernovae and young Galactic supernova
  remnants. The earliest 20 cm luminosity   of J1228+331 (red crosses)
  is placed at an age of 100 years with a horizontal red arrow
  indicating the uncertainty in the age of between 60 and 200 years.
  The earliest luminosity of J1228+441 is probably not the peak 20 cm
  luminosity of J1228+441.  J1228+441 occupies a space between the
  oldest radio supernovae and some of the youngest Galactic supernova
  remnants.   Data for SN 1923A (red x), SN1950B (blue stars), SN
  1957D(grey open squares), SN 1968D (blue filled diamonds), SN 1970G
  (red open circles), SN1978K (blue filled circles), the Crab Nebula
  (green inverted triangle) and Cas A (blue open diamond) were taken
  from (Stockdale et al. 2001, Stockdale et al. 2006, and references
  therein).  Data for SN 1979C (green open triangles), SN   1980K
  (grey filled triangles), and SN 1986J (orange crosses) were taken
  from Stockdale et al. (2006a).  (Stockdale et al. (2006a), New Radio 
Supernova Results,  are
   available online at:
 http://rsd-www.nrl.navy.mil/7213/weiler/sne-home.html and 
 references therein.) \label{rsne}}
\end{figure}

Figure~\ref{rsne} compares  J1228+441  with other core collapse extragalactic radio
 supernovae and young Galactic supernova remnants in a figure adapted from
 Stockdale et al. (2001). In order to compare J1228+441 with other
 supernovae and supernova remnants, the age of J1228+441 was  assumed
 to be 100 years (explosion date 1900 AD, see the discussion of this uncertainty in
 Section~\ref{intro})  and the luminosity was calculated using the
 same model as   Stockdale et al. (2001). The   current luminosity of
 J1228+441  is comparable to the peak luminosity of SN 1970G  and
 higher than the current luminosities  of two Galactic SNRs, Cas A and
 the Crab.  While J1228+441 currently appears in the middle range of
 the radio supernovae luminosities  in Figure~\ref{rsne}, the actual
 peak luminosity of J1228+441 is unknown;  its true peak luminosity
 on Figure~\ref{rsne} could be higher and could be comparable to the
 peak luminosities of SN 	1979C, SN 1986J, and SN 1978K.  It is
 tempting to speculate that J1228+441 occupies a transition region
 between radio supernovae and young Galactic core-collapse SNRs, but this transition region is due mostly to the historical classification of the Crab and Cas A and may not represent a significant difference or transition in the physical properties between SNe and SNRs.

  Cowan, Goss \& Sramak (1991) and  Stockdale et al. (2001) observed
 SN 1970G with the VLA in 1990 and 2000, respectively.  These VLA
 observations of SN 1970G were combined with previous WSRT
 observations  by Gottesman et al. (1972) and Allen et al. (1976) in
 1971 through 1975 so that SN 1970G was observed over  a three decade
 period similar to the time interval  of J1228+441.  Both J1228+441
 and SN 1970G have experienced changes in the spectral index; the
 spectral index of J1228+441  has steepened from $\sim$0.65 in 1996 to
 $\sim$1.0 in  2001-2002 while the spectral index of SN  1970G has
 flattened from 0.56$\pm$0.11 in 1990 to 0.24$\pm$0.20 in 2000
 (Stockdale et al. 2001).  
  
 From the data in Stockdale et al. (2001) and Cowan, Goss, and Sramek  (1991),
 the fractional decline rate for SN 1970G can be calculated for the two periods  with different spectral indices that Stockdale et al. (2001) identified.  Between 1973 and 1991, the decline rate for SN 1970G at 20 cm was $\frac{dS}{S dt}= -6\%\pm
2\%$ per year and between 1991 and 2001, $\frac{dS}{S dt}= -1.1\%\pm
1.3\%$ per year.  The fractional decline rate of J1228+441, $\frac{dS}{S dt}$= -2.8\%$\pm$ 0.6\% per
year at 20 cm, is intermediate to the two values found for SN 1970G.

 Stockdale et al. (2001) suggest that the change in $\alpha$ for SN
1970G could be due to changes in the CSM density.   The radio
supernova, SN 1993J, has also shown a flattening of the spectral index  at
$\sim$ 8 years of age, caused by changes in the deceleration rate
of the shock (Bartel et al. 2002).   Given that J1228+441 is interacting with a
nearby \ion{H}{2} region (Blair, Kirshner, \&  Winkler 1983), the
density profile of the CSM and ISM environment that the remnant shock
is encountering  is likely  to be complex, which could cause future
changes in the deceleration rate of the shock  and the fractional decline rate.  

\section{Summary}

The light curve of J1228+441 shows a significant decline at 6 and 20
cms  with a fractional decline rate of $\frac{dS}{S dt}$= -2.3\%$\pm$
0.4\% per year at 6 cm and  $\frac{dS}{S dt}$= -2.8\%$\pm$ 0.6\% per
year at 20 cm over the  time interval 1975 to 2001.  In the epoch
2001- 2002, a significant steepening of the spectral index was
observed. The spectral index steepened  from $\alpha=0.65 \pm 0.02$
from 1982 to 1996 to $\alpha =1.0 \pm 0.02$ in 2001-2002.  The
spectrum of J1228+441 also steepened in 2001-2002.

 As J1228+441 continues to evolve,  additional monitoring at radio,
 X-ray, and optical wavelengths over multiple epochs is needed in
 order to follow the evolution and determine the properties of this
 young SNR.  Further monitoring  at multiple radio frequencies, with
 the multiple frequencies observed on the same day,   are required to
 better define the light curve. Future changes  in the spectral index
 and the fractional decline rate   could occur if expansion velocities
 and deceleration rates change as the shock encounters changes in the
 density profiles of the CSM and ISM.    Additional optical
 spectroscopy will yield shock velocities, deceleration rates, and
 determine the evolutionary stage and aid in the interpretation of
 the radio observations.
 
\acknowledgements { We thank W. P.  Blair, C. J. Stockdale, and
R. A. Sramek for constructive  comments that helped us improve this
paper. We also thank the referee for a number of helpful comments. }
\section{References}

\parindent 0.0cm

Allen, R. J., Goss, W. M., Ekers, R. D., \& de Bruyn, A. G. 1976,
\aap, 48, 253

Aller, H.D. \& Reynolds, S.P. 1985, \apj, 293L, 73

Anderson, M.C.  \& Rudnick, L. 1995, \apj, 441, 300


Baars, J. W. M., Genzel, R., Pauliny-Toth, I.I.K., \& Witzel, A.,
1977, \aap, 61, 99

Bajaja, E., Huchtmeier, W. K., \& Klein, U. 1994, \aap, 285, 385

Balick, B. \& Heckman, T.  1978, \apjl, 226, L5

Bartel N., Bietenholz, M. F., Rupen, M. P., Beasley, A.J., Graham,
Da.A., Altunun, V.I., Venturi, T., Umana, G., Cannon, W.H.,   \&
Conway, J.E. 2002, \apj, 581, 404

Blair,W.P., Raymond, J.C., Fesen, R.A., \& Gull, T.R. 1984, \apj,
279, 708

Blair, W.P., Kirshner, R. P.,   \& Winkler, P.F., Jr, 1983, \apj,
272, 84

Blair, W. P. 2006, Private communication

Blair, W.P. \& Fesen, R.A., 1998, \baas, 193, 7404

Bignell, R.C. \& Seaquist, E.R., 1983, \apj, 270, 140

Chu, Y.-H., Gruendl, R. A., Stockdale, C. J., Rupen, M. P., Cowan,
J. J.,  \& Teare, S. W., 2004, \aj, 127, 2850 Condon, J. \& Yin, ApJ,
357, 97, 1990

Cowan, J. J., Goss, W. M.,    \& Sramek, R. A., 1991, \apjl, 379L, 49

Cowan, J.J. \& Branch, D., 1982, \apj, 258, 31
 

de Bruyn, A.G., 1983, \aap, 119,301

de Bruyn, A.G., Goss,  W. M., \&, van Woerden, H., 1981, \aap, 94, L25

de Vaucouleurs, G. 1975, in Galaxies and the Universe, ed. A. Sandage,
M. Sandage, \& J. Kristian (Chicago: Univ. Chicago Press), 557




Eck, C. R., Roberts, D. A., Cowan, J. J., \& Branch, D., 1998, \apj,
508, 664

Fabbiano, G., Kim D.-W., \& Trinchieri, G., 1992, \apjs, 80, 531

Goss, W. M.,  Allen, R. J., Ekers, R. D.,  \& de Bruyn, A. G., 1973,
\aap, 243, 42

 Gottesman, S. T., Broderick, J. J., Brown, R. L., Balick, B., \&
 Palmer, P. 1972, \apj, 174, 383

Hunter, D.A., van Woerden, H., \&  Gallagher, J. S., 1999, \aj, 118,
2184

Hunter, D.A., Wilcots, E.M., van Woerden, H., Gallagher, J. S., \&
Kohle, S., 1998, \apj, 495L, 47



Johnson, K. E. et al.,  2006, in preparation.

Kirshner, R. P.  \& Blair, W. P.  1980, \apj, 236, 135

Lacey et al. 2006, in preparation.

Lacey, C. K. \& Duric, N., 2001, \apj, 560, 719

Long, K. S,  Winkler, P. F., \& Blair, W. P., 1992, \apj, 395, 632

 

Montes, M.   J., Weiler, K. W., van Dyk, S.D., Panagia, N., Lacey,
C.K., Sramek, R.A.,   \&  Park, R., 2000, \apj, 532, 1124

Patnaude, D. J.,   \&  Fesen, R.A. 2003, \apj, 587, 221
 
 Pennington, R. L. \& Dufour, R. J., 1983, \apjl, 270L, 7
  
Sandage, A.   \&  Katem, B.  1976, \aj, 81, 743

Seaquist, E.R. \& Bignell, R.C., 1978, \apj, 226, L5

Stockdale, C. J., Goss, W. M., Cowan, J. J., \&  Sramek, R. A., 2001,
\apjl, 559, 139

Stockdale, C. J., Maddox, L.A., Cowan, J. J., Prestwich, A., Kilgard,
R., \&  Immler, S., 2006, \apj, 131, 889.

Stockdale, C. J., Sramek., R. A., Williams, C. L., Van Dyk, S. D.,
Weiler, K. W. , \& Panagia, N., 2006a, New Radio Supernova Results.

Summers,L.K., Stevens, I.R., Strickland, D.K. \& Heckman, T.M., 2003,
\mnras, 342, 690

Thorstensen, Fesen, \& van den Bergh, 2001, \aj, 122, 297

Tuffs, R.J., 1986, \mnras, 219, 13

V\"{o}gler, A. \& Pietsch, W. 1997, \aap, 319, 459

 Weiler, K. W., Panagia, N., Montes, M. J., \& Sramek, R. A., 2002, \araa, 40, 387

\newpage

\begin{deluxetable}{lllcccllll}
\tablecolumns{9}
\tabletypesize\scriptsize
\tablecaption{VLA and WSRT Observations of J1228+441\tablenotemark{a} From 1972 to 2002  \label{taball}}
\tablehead{\colhead{Observing} & \colhead{Telescope} &\colhead{RMS Noise} &\colhead{Beam Size} &\colhead{PA} &\colhead{VLA} &\colhead{Frequency} &\colhead{Flux Density} &\colhead{Error} \\
\colhead{Date} &\colhead{} &\colhead{(mJy/bm)} &\colhead{(\arcsec x \arcsec)} &\colhead{(\degr)} &\colhead{Configuration} &\colhead{(GHz)} &\colhead{(mJy)} &\colhead{(mJy)} }

\startdata
1972 Aug. 01 &WSRT\tablenotemark{b}&1.5&24 x 34&&-&1.420&15.3&2.0 \\
1973 Feb. 22 &  WSRT\tablenotemark{b} &3.0&7 x 10&&-&4.995&13.0&2.0 \\
1973 Nov. 24 &  WSRT\tablenotemark{b} &3.0&7 x 10&&-&4.995& $<$ 8.5 & \\
1974 May 24 &  WSRT\tablenotemark{b} &0.8&7 x 10&&-&4.995&10.9& -2.1, +2.9 \\
1975 Mar. 23 &WSRT\tablenotemark{b} &2.3&56 x 80&&-&0.61&25.5&10 \\
1979 May 02 &  WSRT\tablenotemark{b} &0.4&7 x 10&&-&4.995&8.8& -0.6, +0.8 \\
1979 Oct. 08 & Published VLA\tablenotemark{c} &1.2&1.8 x 1.8&&Mixed&1.479&14.9& 1.2\\
1979 Oct. 08 & Published VLA\tablenotemark{c} &0.15&0.7 x 0.5&&Mixed&4.900&6.8& -0.2, +0.6\\
1980 Jan. 23 &  WSRT\tablenotemark{b} &0.35&7 x 10&&-&4.995&8.6& -0.5, +0.7 \\
1980 Jun. 07 &  WSRT\tablenotemark{b} &1.5&7 x 10&&-&4.874&7.6&1.6 \\
1980 Oct. 22 &  WSRT\tablenotemark{b} &&7 x 10 &&-&4.995&6.4&1.6 \\
1981 Apr. 03 &  WSRT\tablenotemark{b} &&7 x 10 &&-&4.995&7.1&1.0 \\
1981 May 24 &  WSRT\tablenotemark{b} &&7 x 10 &&-&4.995&6.8&1.0 \\
1981 Aug. 05 & Published VLA\tablenotemark{c} &0.19&1.6 x 1.2&&B &4.885&5.6& -0.2, +0.5\\
1981 Oct. 01 & Published VLA\tablenotemark{c} &0.17&15.8 x 11.5&&D &4.885&7.9& 0.2\\
1982 Feb. 02 &Published VLA\tablenotemark{c} &0.3&0.12 x 0.11&&A
&14.965&2.9&-0.3, +0.6\\
1982 Feb. 11 &Published VLA\tablenotemark{c} &0.27&1.0 x 1.0 &&A&1.465&14.7& -0.3, +0.4\\
1982 Feb. 11 & Published VLA\tablenotemark{c} &0.15&0.34 x 0.31&&A &4.885&6.7& -0.2, +0.6 \\
1982 Apr. 11 &  WSRT\tablenotemark{b} &&7 x 10&&-&4.995&6.7&1.0 \\
1982 Jun. 25 &VLA Archive&0.12&1.3 x 1.1&-18&A&1.465&14.7&0.2\\
1982 Jul. 13 &VLA Archive &0.4&4.8 x 3.7& -11&BnA      &1.465&12.6&0.5\\
1982 Jul. 13 & VLA Archive &0.12&1.5 x 1.1 &5 &BnA      &4.885&6.9& 0.1\\ 
1982 Oct. 17 & VLA Archive &0.2&4.1 x 3.7&-21&B    &1.446&13.1& 0.7\\
1983 Mar. 25 & VLA Archive &0.3&15.1 x 12.5&58&C&1.452&15.5& 1.2\\
1983 Mar. 25 & VLA Archive &0.1&4.7 x 3.0&71&C          &4.873&7.3& 0.1\\
1983 Aug. 25 & VLA Archive &0.066&0.38 x 0.34 &-19 &A          &4.873&6.77& 0.07\\
1984 Feb. 10 & VLA Archive &0.27&4.6 x 4.0& 68 &B          &1.490&14.5& 0.35\\
1984 Feb. 10 & VLA Archive &0.08&1.4 x 1.2&62&B          &4.860&5.97& 0.08\\
1984 Dec. 12 &VLA Archive &0.29&1.4 x 1.3&-24&A          &1.490&13.6& 0.3\\
1984 Dec. 12 & VLA Archive &0.14&0.41 x 0.40&-24&A          &4.860&6.4& 0.1\\
1985 May 02 &VLA Archive &0.24&4.5 x 3.7& 55 &B          &1.490&13.2& 0.2\\
1985 May 02 & VLA Archive &0.083&1.4 x 1.1 &56&B          &4.860&5.95& 0.08\\
1986 Dec. 03 &VLA Archive  &0.36&15.1 x 12.7&58&C&1.490&12.4& 0.8\\
1986 Dec. 03 &VLA Archive  &0.06&4.5 x 4.1 &-86&C&4.860&6.02& 0.07\\
1987 Oct. 15 &VLA Archive  &0.21&3.5 x 1.4&-60&BnA   &1.490&11.3& 0.3\\
1987 Oct. 15 & VLA Archive &0.075&1.2 x 0.39&-61&BnA   &4.860&6.09& 0.08\\
1988 Jan. 17 & VLA Archive &0.14&1.3 x 1.3 &-61&B          &4.873&5.0 & 0.15\\
1988 Feb. 12 &VLA Archive  &0.18&3.9 x 1.6&75&CnB&4.873&5.6&0.2\\
1988 May 26 & VLA Archive &0.065&9.0 x 4.4&77& DnC&4.873&5.35& 0.4\\
1988 Dec. 30 &VLA Archive  &0.22&1.6 x 1.2& 81 &A          &1.490&12.7& 0.5\\
1989 Mar. 13 & VLA Archive &0.088 &1.3 x 1.2 &-71& B
&4.860&5.62& 0.11\\
1990 Jul. 29 & VLA Archive &0.24&1.4 x 1.2 &45&B &4.860&5.9& 0.3\\
1991 Dec. 06 & VLA Archive &0.12&1.3 x 1.2 &-33&B &4.860&4.9& 0.15\\
1994 Mar. 30 &VLA this paper  &0.026&1.3 x 1.2 & 86 &A&1.460&9.55& 0.03 \\
1994 May 07 & VLA Archive &0.25&0.71 x 0.41&-52&A$\rightarrow$B &4.873&5.4& 0.5\\
1994 Jun. 16 &VLA this paper &0.028&2.42 x 1.39&82&B &4.860&4.22& 0.03 \\
1994 Sep. 05 & VLA Archive &0.18&4.8 x 4.1&71&B &1.420&10.0 & 0.3\\
1996 Sep. 01 &VLA Archive &0.016&13.4 x 12.1&83&D          &4.860&4.71& 0.2\\
1996 Sep. 01 &VLA Archive&0.015& 8.3 x 7.1 & -88 &D &8.460&3.25&0.05\\
1996 Oct. 18 &VLA Archive&0.22&0.19 x 0.19& -45 &A &14.940&2.3&0.5\\
1998 Sep. 14 &VLA Archive&0.093&3.9 x 3.5&73&B&1.719&7.5&0.09\\
2001\tablenotemark{d}  &VLA\tablenotemark{e} &0.011&0.86 x 0.81&53&
B,B$\rightarrow$C,C
&8.460&2.53&0.01\\
2002 Feb. 16 &VLA\tablenotemark{e} &0.04&0.42 x 0.36&-89&A&4.860&4.0 & 0.06 \\
2002 Nov. 15 &VLA\tablenotemark{e} &0.039&1.03 x 0.08&78& C
&22.460&0.86&0.07\\
\enddata
\tablenotetext{a}{The location of the SNR was measured from a 2001 high resolution 8.4 GHz 
 image (Johnson et al. 2006) that provided  the best measure of the SNR's position:  $\alpha (2000): 12^h 28^m 10\fs95 \pm 0\fs02,   \delta(2000):
44\arcdeg 06\arcmin 48\farcs6 \pm 0\farcs2$.}
\tablenotetext{b}{De Bruyn, 1983, \aap,  119, 301}  
\tablenotetext{c}{Bignell et al., 1983, \apj, 270, 140}
 \tablenotetext{d}{2001 Apr. 7, 2001 May 17, 2001 Jun 29, 2001 Sep. 1}
\tablenotetext{e}{Johnson et al., 2006, in preparation}
\end{deluxetable}



\begin{deluxetable}{llcccc}
\tablecaption{The Spectral Index of J1228+441 from 1972 to 2002
\label{alpha}} 
\tablehead{ \colhead{Date$_1$}  &\colhead{ Date$_2$}
& \colhead{$\nu_1$ (GHz)} &  \colhead{$\nu_2$ (GHz)} &
\colhead{$\alpha$} & \colhead{Error} }
\startdata 
1975 Mar 23 &  1974 May 24  &   0.610 & 4.995 &        0.40 &   0.23 \\
1979 Oct 8 &   1979 May 2 &     1.479 & 4.885 &        0.44 &   0.10 \\
1979 Oct 8 &   1979 Oct 8 &     1.479 & 4.900 &        0.65 &   0.10 \\
1979  Oct 8  & 1979 May 2  &    1.479 & 4.995 &        0.43 &   0.10 \\
1982  Feb 11  &1982  Feb 11  &  1.465 & 4.885 &        0.65 &   0.08 \\
1982 Feb 11 &  1982 Feb 23  &   4.885 & 14.965 &       0.75 &   0.20 \\
1982 Jul 13 &  1982 Jul 13 &    1.465 & 4.885 &        0.50 &   0.04 \\
1984  Feb 10 & 1984 Feb 10 &    1.490 & 4.860 &        0.75 &   0.02 \\
1984 Dec 12 &  1984 Dec 12 &    1.490 & 4.860 &        0.64 &   0.02 \\
1985 May 2  &  1985 May 2  &    1.490 & 4.860 &        0.67 &   0.02 \\
1986 Dec 3 &   1986 Dec 3 &     1.490 & 4.860 &        0.61 &   0.06 \\
1987 Oct 15 &  1987 Oct 15 &    1.490 & 4.860 &        0.52 &   0.03 \\
1988 Dec 30 &  1988 May 26 &    1.490 & 4.873 &        0.73 &   0.07 \\
1988 Dec 30  & 1988 Feb 12 &    1.490 & 4.873 &        0.69 &   0.04 \\
1988 Dec 30  & 1988 Jan 17 &    1.490 & 4.873 &        0.79 &   0.04 \\
1988 Dec 30 &  1989 Mar 13 &    1.490 & 4.873 &        0.69 &   0.04 \\
1994 Mar 30 & 1994 May 7 &     1.460 & 4.873 &        0.47 &   0.08 \\
1994 Sep 5 &   1994 Jun 16 &    1.420 & 4.860 &        0.70 &   0.03 \\
1996 Sep 1 &   1996 Oct 18 &    4.860 & 14.940 &       0.64 &   0.20 \\
1996 Sep 1  &  1996 Oct 18 &    8.460 & 14.940 &       0.61 &   0.38 \\
2002 Feb 16 &  2001 Jun 29 &    4.860 & 8.460 &        0.83 &   0.03 \\
2002 Feb 16 &  2002 Nov 15 &    4.860 & 22.460 &       1.00 &   0.05 \\
2001 Jun 29 &  2002 Nov 15 &    8.460 & 22.460 &       1.11 &   0.08 \\
\enddata
\end{deluxetable} 




\begin{deluxetable}{lllll}
 
\tablecaption{Comparison of J1228+441 with Old Radio
Supernovae\tablenotemark{a}
\label{radiosne}} 
\tablehead{ \colhead{Parameter}  & \colhead{SN 1970G}  &  \colhead{SN
 1957D} &  \colhead{SN 1950B} & \colhead{J1228+441 }}

\startdata  
 Distance
(Mpc)	&7.4	&4.1&	4.1 &  3.9\\
Peak 20
cm Flux density (mJy)	&5.9 $\pm$
1.6 &	2.7 $\pm$ 0.12&	0.7$\pm$ 0.08& 14.9  $\pm$
1.2\tablenotemark{b}\\
~~~ Supernova age (yr)&	19.74&	25.24&	33.00 &
$\sim$100\\
~~~ Luminosity (ergs
s$^{-1}$ Hz$^{-1}$)	&4.0$\times 10^{26}$ &5.3$\times 10^{25}$
	&1.4$\times 10^{25}$ &  2.8$\times 10^{26}$\\
Peak 6 cm Flux density
(mJy)	&0.12 $\pm$ 0.020	&1.39 $\pm$ 0.04	&0.37 $\pm$
0.03 &4.0 $\pm$0.06\tablenotemark{b}\\
~~~ Supernova age (yr)	&30.50& 32.82&	40.58
& $\sim$100 \\
~~~ Luminosity  (ergs s$^{-1}$ Hz$^{-1}$)
&7.9 $\times 10^{24}$&	2.8 $\times 10^{25}$&	7.4 $\times 10^{24} $
& 7.6 $\times 10^{25} $ \\ 
Most recent spectral index
$\alpha$	&0.24 $\pm$ 0.20&	0.11 $\pm$ 0.06&	0.57
$\pm$ 0.08 & 1.0 $\pm$ 0.02\tablenotemark{c}\\
Earlier spectral index
$\alpha$&	0.56 $\pm$ 0.11\tablenotemark{d}&	0.23 $\pm$ 0.04	&0.55
$\pm$ 0.13 &0.65 $\pm$ 0.02\tablenotemark{e}\\
 \enddata

\tablenotetext{a}{Table adapted from Stockdale et al. (2001). Values
 for SN 1970G, SN 1957D, and SN 1950 B were taken from Stockdale et
 al. (2001) and references therein. } 
 \tablenotetext{b}{ Peak flux density for J1228+441 is not known.  The current maximum observed flux density is quoted here.}
 \tablenotetext{c}{ $S\propto \nu^{-\alpha}$, $\alpha$ based on 22.4 , 8.46, and 4.86 GHz
 observations obtained in 2001-2002.}
\tablenotetext{d}{ $\alpha$ based on 1.4 GHz and 8.5 GHz  observations  (Stockdale et al. 2001).}
\tablenotetext{e}{Average  $\alpha$ from 1982 through 1997.}

\end{deluxetable}


\end{document}